\documentclass[twocolumn,aps,showpacs,preprintnumbers,a4paper]{revtex4}
\usepackage{graphicx}
\usepackage{bm}
\usepackage{amssymb}
\usepackage{subfigure}
\begin{document}
\preprint{v20111226}

\title{Surface properties of the clean and Au/Pd covered Fe$_3$O$_4$(111):\\ a DFT and DFT+$U$ study}

\author{Adam Kiejna}\email{kiejna@ifd.uni.wroc.pl} 
\author{Tomasz Ossowski} 
\author{Tomasz Pabisiak}

\affiliation{Institute of Experimental Physics, University of Wroc{\l}aw, Plac M. Borna 9, PL-50-204 Wroc{\l}aw, Poland}


\begin{abstract}
The spin-density functional theory (DFT) and DFT+$U$ with Hubbard $U$ term accounting for on-site Coulomb interactions were applied to investigate structure, stability, and electronic properties of different terminations of the Fe$_3$O$_4$(111) surface.  All terminations of the ferrimagnetic Fe$_3$O$_4$(111) surface exhibit very large (up to 90\%) relaxations of the first four interlayer distances, decreasing with the oxide layer depth. Our calculations predict the iron terminated surface to be most stable in a wide range of the accessible values of the oxygen chemical potential. The adsorption of Au and Pd on two stable Fe- and O-terminated surfaces is studied. Our results show that Pd binds stronger than Au both to the Fe- and O-terminated surface. DFT+$U$ gives stronger bonding than DFT. The bonding of both adsorbates to the O-terminated magnetite surface is by 1.5-2.5 eV stronger than to the Fe-terminated surface.
\end{abstract}


\maketitle

\section{Introduction}

Magnetite (Fe$_3$O$_4$), the earliest known magnetic material, is an abundant mineral in nature and is important due to its potential applications in magnetoelectronics and in catalysis. It is one of the most interesting iron oxides formed during the corrosion processes (rusting). Ultra-thin films and nanostructures formed by noble metals on iron-oxide support show enhanced catalytic properties compared with a clean oxide surfaces. Thus, understanding the properties of magnetite surfaces is of utmost importance from the viewpoint of basic science and applications. However, even clean iron oxide surfaces are relatively little explored which is connected with difficulties in the preparation of well defined surfaces \cite{WeiRan02,GonFN08}.

At room temperatures and under normal pressure conditions magnetite crystallizes in the inverse spinel structure (space group $Fd\bar{3}m$) in which tetrahedral positions are occupied by ferric (Fe$^{3+}$)  while octahedral  ones contain equal number of ferric and ferrous (Fe$^{2+}$) iron atoms. Magnetite's  primitive rhombohedral cell contains two formula units of Fe$_3$O$_4$ and its volume is equal to one quarter of the spinel unit cell. Bulk magnetite is a semi-metal. Since the Fe$^{3+}$ ions are aligned antiferromagnetically and the ratio of  Fe$^{3+}$ and Fe$^{2+}$ ions is 2:1 the overall crystal structure is ferrimagnetic. 
At $T\approx 120$ K, the metal-insulator Verwey transition occurs which is connected with a long-range change of the degree of localization of electrons in the octahedral Fe atoms \cite{PiePO96,Lodz07,RowPG09}. 

The structure of magnetite can be also represented by the hexagonal conventional unit cell which contains eight formula units. In this stacking, (111) oriented layers of oxygen atoms separate alternate Fe monolayers of octahedral (Fe$_{\rm oct1}$) sites  and Fe trilayers consisting of a Fe$_{\rm oct2}$ monolayer in octahedral sites with Fe$_{\rm tet1}$ and Fe$_{\rm tet2}$ layers in tetrahedral sites on either side.  Thus the stacking sequence of the atomic planes perpendicular to the [111] direction can be written \cite{WeiRan02} as Fe$_{\rm oct2}$-Fe$_{\rm tet1}$-O$_1$-Fe$_{\rm oct1}$-O$_2$-Fe$_{\rm tet2}$-. 
The (111) surface, which is the dominant cleavage plane of magnetite  and is often exposed on naturally grown crystals can have six different terminations \cite{WeiRan02}. 
Only four of them have been confirmed experimentally, namely, Fe$_{\rm oct2}$ \cite{LenCLMTV96}, Fe$_{\rm tet1}$ \cite{RitWei99}, and densely packed oxygen planes O$_2$, and O$_1$ \cite{BerMMS04,BerMMS04a}. 
Both iron and oxygen terminated Fe$_3$O$_4$(111) surfaces are polar of type III  according to Tasker's classification \cite{Tasker79}. The surface termination is very sensitive to the preparation conditions of the samples \cite{WeiRan02} which are either Fe$_3$O$_4$ single crystals or epitaxial Fe$_3$O$_4$ films on single crystal substrates. Despite many experimental investigations, information about the (111) surface at the atomic level is very scarce. There are not many data about its magnetic properties or how its structure and composition change with temperature and oxygen pressure. 
Magnetite is a strongly correlated system and DFT with standard (local or semi-local) exchange-correlation functionals does not allow for a correct description of its electronic structure because of inadequate treatment of the strong Coulomb interaction between $3d$ electrons localized on the Fe ions. 
This shortcoming of DFT is corrected in practice by either of two semi-empirical approaches: hybrid functionals, where the exact Hartree-Fock exchange is partially mixed with the DFT exchange, or DFT+$U$, where the on-site Coulomb repulsion is described by an additional Hubbard term $U$. 

The question what is the most stable termination of the Fe$_3$O$_4$(111) surface is still open and controversies about its structural details seem to remain. An earlier scanning tunneling microscopy (STM) study \cite{LenCLMTV96} reported two coexisting surface terminations. One was assigned to Fe$_{\rm oct1}$ atoms and the other to Fe$_{\rm oct2}$--Fe$_{\rm tet1}$ layers. A low energy electron diffraction (LEED) study \cite{RitWei99} concluded that the (111) surface termination corresponds to 1/4 monolayer of Fe atoms over a hexagonal close-packed O layer underneath. This disagreed with \textit{ab initio} periodic Hartree-Fock calculations of Ahdjoudj \textit{et al.} \cite{AhdMMvHS99} who found the Fe$_{\rm oct2}$--Fe$_{\rm tet1}$ bilayer to be the most favorable termination of the clean surface. Lemire \textit{et al.}~\cite{LemMHSF04} studied the surface structure of Fe$_3$O$_4$(111) films by CO adsorption and concluded that the (111) surface is terminated with Fe$_{\rm oct2}$. 
Recent full-potential linearized augmented plane-wave (FP-LAPW) calculations by Zhu \textit{et al.} \cite{ZhuYL06} determined the structure, composition and relative stability of five different terminations of the Fe$_3$O$_4$(111) surface. The effect of different functionals for  exchange and correlation energy (generalized gradient approximation (GGA), local density approximation+$U$ (LDA+$U$)) has been also discussed. According to that study the Fe$_{\rm oct2}$ termination is the most stable one. This, however, was not confirmed by a more recent first principles calculations performed by Grillo \textit{et al.} \cite{GriFR08} within GGA+$U$ and Martin \textit{et al.} \cite{MarCVW09} using GGA, according to which the Fe$_{\rm tet1}$-terminated surface has the lowest surface energy. Recent STM experiments have reported that the stoichiometric (111) surface of magnetite corresponds to Fe$_{\rm tet1}$ termination \cite{PauSCSB07}. Another, recent combined STM and first principles calculation study \cite{ShiJKKK10} also predicted the Fe$_{\rm tet1}$ termination as the most stable one. 

The adsorption of metal atoms on magnetite was very rarely studied. On the theoretical side only the interaction of alkali metal atoms with the Fe$_{\rm tet1}$-terminated Fe$_3$O$_4$(111) surface has been studied using DFT at the GGA level \cite{YanWLWJ09}. To our knowledge, studies of the adsorption of noble and transition metal atoms on Fe$_3$O$_4$(111), which are important in catalysis, have not been reported so far.  

In this work we revisit first the calculations for different terminations of the clean Fe$_3$O$_4$(111) surface using DFT and DFT+$U$ approaches, in order to explore the influence of strong on-site electronic correlations on the structure and physical properties of magnetite, and to form a firm basis for our studies of Au and Pd atom adsorption on the Fe$_3$O$_4$(111) surface, which is the main subject of this work.

\begin{figure}
\includegraphics*[width=8.4cm]{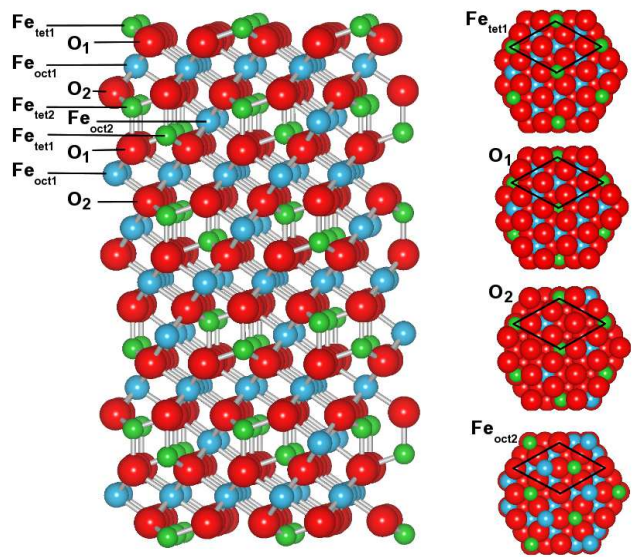}
\caption{(Color online) The Fe$_3$O$_4(111)$ slab used in the surface calculations. The successive terminations were created by removing top and bottom layer from the thickest Fe$_{\rm tet1}$ terminated slab. Iron and oxygen atoms are represented by small and large balls respectively. The right hand side figures show top views of the considered terminations. The parallelograms mark the 1$\times$1 surface cell applied in the calculations.}\label{fig1_slab}
\end{figure} 

\section{Methods and computational details}

The calculations presented in this work are based on the spin density functional theory as implemented in the VASP package \cite{KreHaf93,KreFur96}. The calculations employed the GGA-PW91 version of the exchange and correlation energy functional \cite{PerCVJPSF92} with the spin interpolation of Vosko \textit{et al.} \cite{VosWN80}, and the GGA plus on-site Coulomb interaction term $U$ (GGA+$U$) using the Dudarev \textit{et al.} \cite{DudBSHS98} approach. Following previous calculations \cite{JenGH06,Lodz07} for the bulk and surfaces of Fe$_3$O$_4$ the GGA+$U$ calculations were performed with the effective parameter of interaction between electrons $U_{\rm eff}=U-J=3.61$ eV (the Coulomb and screened exchange parameters  $(U,J)=(4.5,0.89)$ eV, respectively). 
The electron ion-core interactions were described by the  projector augmented wave (PAW) method \cite{KreJou98}. The plane waves basis with cut-off energy of 500 eV and the conjugate gradient algorithm were applied to determine the electronic ground state. The integrations over the Brillouin zone were performed using Monkhorst-Pack grids \cite{MonPac76}. A Gaussian broadening of the Fermi surface of 0.2 eV was applied to improve the convergence of the solutions. The results presented in this work were obtained using $k$-point meshes of 6$\times$6$\times$6 for the bulk, and 6$\times$6$\times$1 for surface calculations, which allowed to obtain total energy convergence within 1 meV. For surface calculations $\Gamma$-centered grids were used.

\begin{figure*}
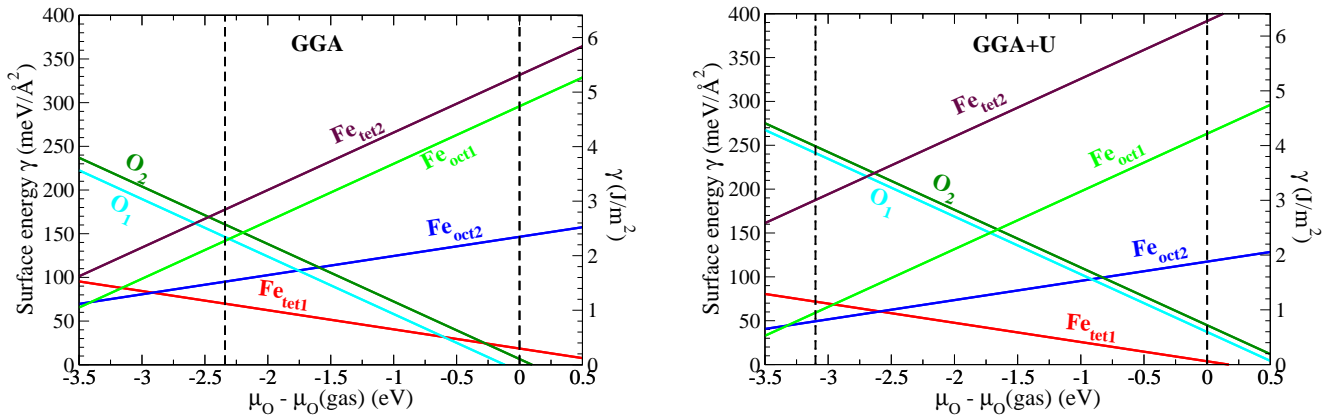

\subfigure{\includegraphics*[width=8.3cm]{./fig_2a.eps}}
\hspace{0.5cm}
\subfigure{\includegraphics*[width=8.3cm]{./fig_2b.eps}}
\caption{(Color online) The dependence of surface energy on the oxygen chemical potential, $\mu_{\rm O}$, showing relative stability of six terminations of the Fe$_3$O$_4(111)$ surface calculated within the GGA and GGA+$U$. Vertical, dashed lines mark allowed range of the oxygen chemical potential.  \label{phsdiag}
} 
\end{figure*}

Calculations of the Fe$_3$O$_4$ bulk structure showed that among the nonmagnetic and different magnetic phases, the ferrimagnetic phase is most stable, with the magnetic moments on the Fe$_{\rm oct}$ atoms antiparallel to those on the Fe$_{\rm tet}$ atoms. The lattice constant and bulk modulus of the magnetite crystal calculated within GGA,  8.377 {\AA}  and 172 GPa, respectively are in very good agreement with experiment (8.396 {\AA} \cite{OkuKM96}, 8.393 \AA\ \cite{Fleet82}; 181 GPa \cite{OkuKM96}) and other GGA calculations \cite{MarCVW09,PinEll06}. The GGA magnetic moments on the Fe$_{\rm tet}$ and Fe$_{\rm oct}$ atoms are $-$3.45$\mu_{B}$ and 3.49-3.61$\mu_{B}$ respectively. The moments on the O atoms are much smaller ($\sim$0.08$\mu_B$). The total magnetic moment (3.66$\mu_{B}$) per formula unit, is about 0.4$\mu_{B}$ lower than the experimental value \cite{Aragon92}. The GGA+$U$ calculations give the lattice constant (8.473 {\AA}) and bulk modulus (182 GPa) in good agreement with experimental data and other calculations \cite{PinEll06,GriFR08}. They improve the magnetic moments on the Fe atoms compared to GGA and give 4.04$\mu_{B}$, and 3.91-3.95$\mu_{B}$ on the Fe$_{\rm tet}$ and Fe$_{\rm oct}$ atoms, respectively. With the moments on the O atoms reduced by 50\%, the total magnetic moment per formula unit is 3.97$\mu_{B}$, in very good agreement with experiment (4.05$\mu_B$ \cite{Aragon92}).

The optimized lattice parameters were used to construct the Fe$_3$O$_4$(111) surface slabs consisting of 19 to 29 atomic layers separated by a vacuum region. Starting from the thickest symmetric slab for the Fe$_{\rm tet1}$-termination (Fig.~\ref{fig1_slab}) the other terminations were created by striping off subsequent atomic layers, without changing the supercell size, both from the top and the bottom of the slab. Thus the surface was separated from its periodic replicas by vacuum region ranging from 15.5 to 24 {\AA} (the latter is for the thinnest Fe$_{\rm oct2}$-terminated slab). In surface calculations the positions of all atoms were relaxed until the forces were smaller than 0.02 eV/\AA.

The stability of different terminations of the Fe$_3$O$_4(111)$ surface as a function of oxygen partial pressure was considered based on the \textit{ab initio} thermodynamics \cite{ReuSch01} in which the surface free energy $\gamma (T,P) $ is expressed  as a function of the pressure and temperature by the chemical potentials $\mu_{\rm Fe}$, $\mu_{\rm O}$ of the constituents: 
\begin{equation}
\gamma(T,P) = {1\over{2A}}\left[G_{{\rm Fe}_3{\rm O}_4}^{\rm slab} - N_{\rm Fe} \mu_{\rm Fe}(T,P) -N_{\rm O}\mu_{\rm O}(T,P)\right],  \label{eq3}
\end{equation}
where the Gibbs free energy, $G_{{\rm Fe}_3{\rm O}_4}^{\rm slab}$, can be expressed by the total energy of the slab $E_{\rm tot}^{\rm slab}$, and $N_{\rm Fe}$ and $N_{\rm O}$ represent the number of Fe and O atoms in the system. The Gibbs free energy per magnetite formula unit, $g$, is related to the chemical potentials of iron and oxygen  through the relation $g^{\rm bulk}_{{\rm Fe}_3{\rm O}_4}= 3\mu_{\rm Fe} + 4\mu_{\rm O}$. Consequently, the surface energy as a function of oxygen chemical potential can be written as
\begin{equation}
\gamma = {1\over{2A}}\left[E_{\rm tot}^{\rm slab} - {1\over 3}N_{\rm Fe} g^{\rm bulk}_{{\rm Fe}_3{\rm O}_4} + \left({4\over 3}N_{\rm Fe} - N_{\rm O}\right)\mu_{\rm O}\right], \label{eq6}
\end{equation}
where the Gibbs free energy is approximated by the internal energy from DFT calculations \cite{ReuSch01}. $\mu_{\rm O}$ is referenced with respect to the chemical potential of oxygen in a gas phase $\mu^{\rm gas}_{\rm O} = {{1}\over{2}}E^{\rm tot}_{{\rm O}_2}$, where the total energy of an oxygen molecule $E^{\rm tot}_{{\rm O}_2}$ is calculated in a large box. 

Au and Pd atoms were adsorbed in four different adsorption sites of the 1$\times$1 surface unit cell, on both sides of the relaxed, symmetric  Fe$_3$O$_4$(111) slab. The adsorption binding energy was calculated as 
\begin{equation}
     E_{\rm ad} = -(E^{{\rm X}/{\rm sub}} -E^{\rm sub} - 2E^{\rm X})/2 ,  
\end{equation} 
where $E^{{\rm X/sub}}$ is the total energy of the slab covered with adsorbate X, $E^{\rm sub}$ represents the energy of the relaxed bare oxide support, and $E^{\rm X}$ is the energy of a free adsorbate atom. 

\section{Results}

\subsection{Clean Fe$_3$O$_4$(111) surface}

The variation of surface energy as a function of the oxygen chemical potential $\mu_{\rm O}$ is displayed in Fig.~\ref{phsdiag}. The accessible range of $\mu_{\rm O}$  is limited, at the lower limit by the onset of magnetite decomposition to form bulk iron and, at the upper limit, by start of oxygen condensation on the magnetite surface. It is seen that surface composition of the Fe$_{\rm tet1}$ termination minimizes the surface free energy over a wide range of $\mu_{\rm O}$ for both types of calculations, in agreement with previous predictions \cite{GriFR08,PauSCSB07,ShiJKKK10}. 
However, at low $\mu_{\rm O}$ which correspond to very low oxygen pressures the GGA+$U$ results show that the Fe$_{\rm oct2}$ termination may turn out to be stable. Zhu et al. \cite{ZhuYL06} have predicted this termination to be most stable. The GGA+$U$ does not give the stable oxygen terminations of the Fe$_3$O$_4$(111) surface seen at higher pressures in the GGA results. In the following we investigated only four clean surfaces of magnetite -- the two iron terminated (Fe$_{\rm tet1}$, Fe$_{\rm oct2}$), and two oxygen terminated (O$_1$ and O$_2$) ones.  

\begin{figure}
\includegraphics[width=6.5cm]{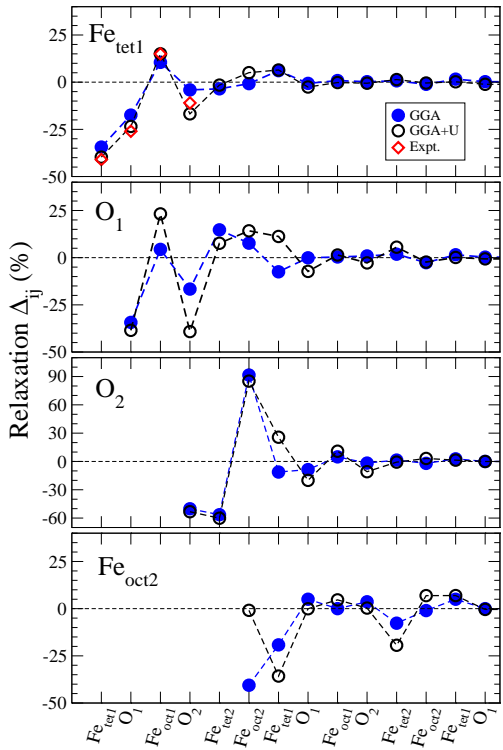}
\caption{(Color online) Relaxations, $\Delta_{ij}$, of the interplanar distance for different Fe$_3$O$_4$(111) terminations. Experimental data for the Fe$_{\rm tet1}$ termination are taken from Ref. \onlinecite{RitWei99}. In the ideal crystal bulk the subsequent separations of the (111) layers calculated within GGA (GGA+$U$) are as follows: Fe$_{\rm tet1}$-O$_1$, 0.637 (0.644) \AA; O$_1$-Fe$_{\rm oct1}$ and Fe$_{\rm oct1}$-O$_2$, 1.176 (1.190) \AA; O$_2$-Fe$_{\rm tet2}$, 0.637 (0.644) \AA; Fe$_{\rm tet2}$-Fe$_{\rm oct2}$ and Fe$_{\rm oct2}$-Fe$_{\rm tet1}$, 0.605 (0.611) \AA.
} \label{relaxations}
\end{figure}

After structural optimization the surfaces are strongly relaxed. The relaxation of the interplanar distance is calculated as $\Delta_{ij}=(d_{ij}-d)/d$, where $d_{ij}$ is the distance of the relaxed $i$ and $j=i+1$ planes, and $d$ is the corresponding distance in the bulk. The calculated relaxations (Fig.~\ref{relaxations}) show an oscillatory character, though not very regular, of the contraction-expansion type. In general, the geometry changes calculated within the GGA and GGA+$U$ approaches are qualitatively similar and predict a large contraction of the first interplanar distance. They differ, however, in the topmost layer relaxation of the Fe$_{\rm oct2}$ termination, where GGA predicts a large (40\%) contraction of the Fe$_{\rm oct2}$-Fe$_{\rm tet1}$-layer distance, whereas GGA+$U$ shows no relaxation. Top layers relaxation calculated within LDA+$U$ \cite{ZhuYL06} are substantially larger compared with our GGA values. 
As seen from  Fig.~\ref{relaxations} for the Fe$_{\rm tet1}$ terminated surface the relaxations are in very good agreement with experimental data \cite{RitWei99}, and agree well with results of the LDA+$U$ calculations \cite{ZhuYL06}. The largest relaxations (up to about 90\%) are observed in the O$_2$-termination.  The first and second interlayer distances exhibit large contractions of about 50\% GGA (53\%, GGA+$U$) and 56\% GGA (60\%, GGA+$U$), respectively. These are almost compensated by a large expansion of the third interlayer spacing (92\%, GGA; 85\%, GGA+$U$). The relaxations of the O$_2$ surface are much larger (up to three times) than those found from LDA+$U$ \cite{ZhuYL06}.
They lead to a specific final configuration: a surface region of reduced thickness formed by a triple-layer, slightly separated from the deeper lying oxide layers. Similar structures are observed for the other considered terminations: separated triple layers for the Fe$_{\rm tet1}$- and Fe$_{\rm oct2}$- terminations, and a separated double layer for the O$_1$-terminated surface. In, general, the relaxations of the O-terminated surfaces of magnetite are much larger than those observed on O-terminated hematite (0001) surface \cite{KiePab11}. The topmost layer relaxations on the iron terminated surface of magnetite (111) show similarity to those calculated for hematite (0001) surfaces \cite{KiePab11}. 
The calculated contraction of the spacing between the topmost Fe and O surface layers might be considered as a mechanism which reduces the total electrostatic dipole moment and thus stabilizes these polar surfaces. The ideal bulk O-layers show only a small corrugation. On the O$_1$ plane one of the O atoms is slightly shifted down giving a 0.03 \AA\ corrugation with respect to the average position of the plane. A similar corrugation is caused by an upward shift of one of the O atoms of the O$_2$ plane. 

\begin{figure*}
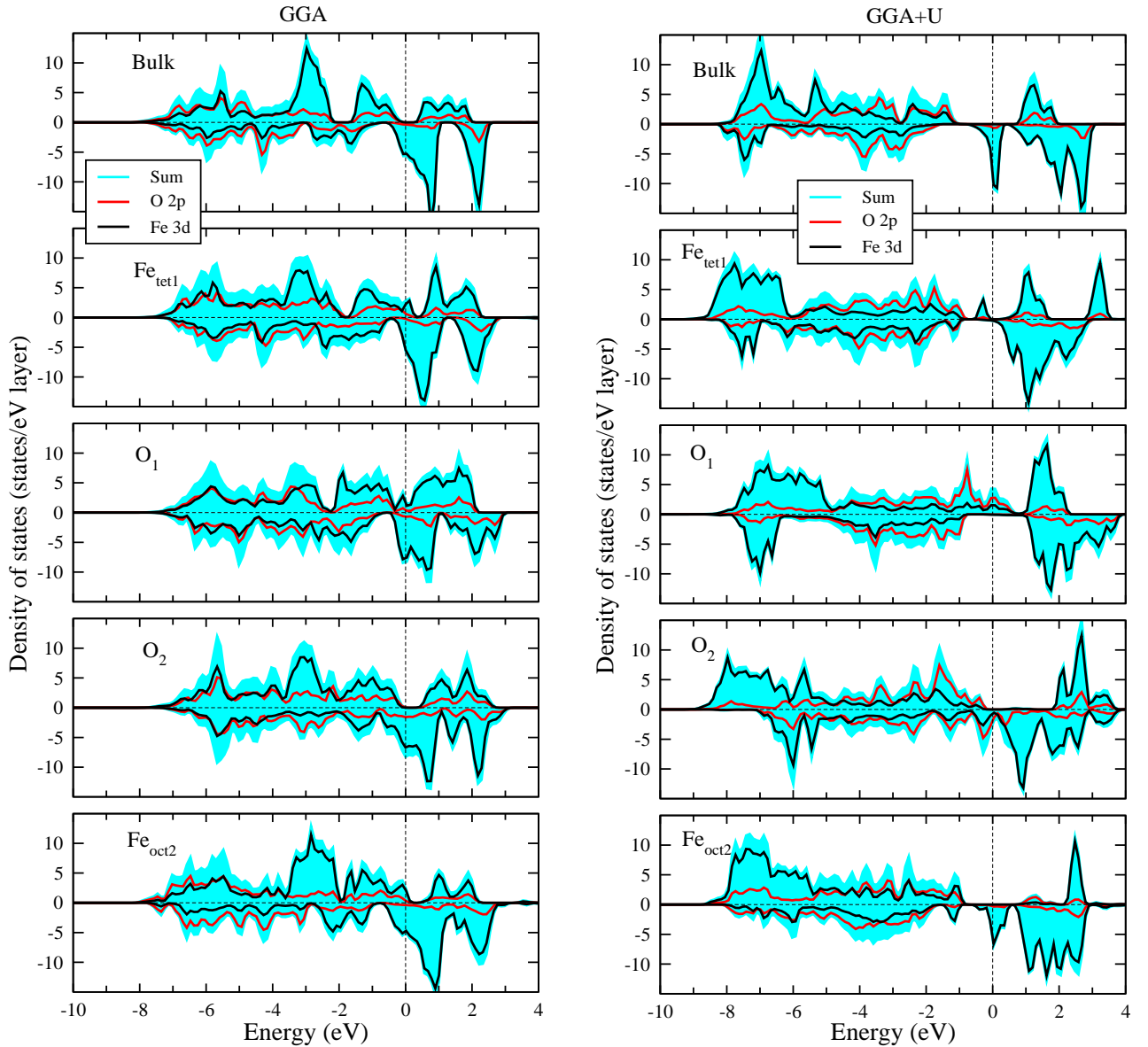

\subfigure{\includegraphics*[width=8.0cm]{./fig_4a.eps}}
\hspace{0.5cm}
\subfigure{\includegraphics*[width=8.0cm]{./fig_4b.eps}}
\caption{(Color online) Density of states of bulk magnetite and LDOS on atoms of six topmost atomic layers of different surface terminations decomposed into the contributions from the iron and oxygen atomic layers. Majority and minority spin states are respectively displayed as positive and negative. The left and right hand side panels show respectively GGA and GGA+$U$ results. } \label{f4_DOS_all}
\end{figure*} 

Figure \ref{f4_DOS_all} shows the calculated density of states (DOS) of bulk magnetite and the local density of states (LDOS) at the four Fe$_3$O$_4$(111) surface terminations. As is seen the DOS and LDOS resulting from the respective GGA and GGA+$U$ calculations differ substantially. Both methods predict half-metallic behavior \cite{ZhaSap91} of the bulk magnetite with the band gap in majority spin electrons which is much wider when calculated within GGA+$U$. However, all considered (but O$_2$) surface terminations show metallic LDOS when calculated within GGA. The metallic character of the O$_1$ termination agrees with the results of a combined spin-polarized STM and DFT study reported by Berdunov \textit{et al.} \cite{BerMMS04a}. This prediction is at variance with the results from photoemission spectroscopy studies \cite{SchSTFKSSMBC05,FonDPRG07} and with our GGA+$U$ results (Fig. \ref{f4_DOS_all}) which show that almost all terminations remain half-metallic. The exception is Fe$_{\rm tet1}$ surface where the narrow ($\sim$0.3 eV) energy gap opens below the Fermi level. 
The location of the band gap alters from the majority to minority spin band, depending on the termination. Compared with the GGA the LDOS calculated within GGA+$U$ extend over a wider energy range and the main weights of the Fe$_{\rm tet1}$ and Fe$_{\rm oct2}$ states are shifted to lower energies. At the Fe$_{\rm oct2}$ termination responsible for the half-metallic character are Fe $3d$ states, whereas at the O$_1$ and O$_2 $ terminations half-metallicity is due to the hybridized O $2p$ and Fe $3d$ states. 

\begin{table}
\caption{Work function of different Fe$_3$O$_4(111)$ surfaces. \label{tab1-WF} }
\begin{ruledtabular}
\begin{tabular}{lcc}
Termination     & \multicolumn{2}{c}{Work function (eV)} \\ \cline{2-3}
 & GGA & GGA+$U$\\ \hline
Fe$_{\rm tet1}$  & 4.93  & 5.48 \\
Fe$_{\rm oct2}$  & 3.58  & 3.90 \\
O$_1$            & 7.04  & 8.09 \\
O$_2$		 & 6.88  & 7.66 
\end{tabular}
\end{ruledtabular}
\end{table}
The difference in the electronic structure of the differently terminated surfaces is manifested in their work function values. It was  calculated as the difference between the electrostatic potential in the vacuum and the Fermi energy of the slab. The work function of the O-terminated surfaces is 2-3.5 eV (GGA) and 2-4 eV (GGA+$U$) larger than that of the Fe-terminated surfaces. The GGA work functions are about 0.4-1.0 eV lower than those determined from GGA+$U$ (Table \ref{tab1-WF}). Generally, the work function is lowest for the Fe$_{\rm oct2}$-terminated surface and highest for the O$_1$ termination. 
The calculated Bader charges \cite{Bader90,HenAJ06} on atoms at the exposed surface terminations show that the atoms in the topmost Fe layer, both of the Fe$_{\rm tet1}$ and Fe$_{\rm oct2}$-terminated surfaces, gain electron charge compared to that on atoms in the bulk crystal layers (0.28$e$; 0.76$e$ and 0.27$e$, respectively). The oxygen atoms of the O-terminated surfaces, as well as the O-atoms of the Fe$_{\rm tet1}$ termination, lose electrons. In contrast, in the Fe$_{\rm oct2}$ termination, not only the Fe$_{\rm oct2}$ and  Fe$_{\rm tet1}$ atoms gain electrons but also most of the O-atoms of the subsurface layers except for one of the atoms of the O$_1$-layer which loses 0.14$e$.

The changes in the magnetic moments on the iron atoms due to the presence of the surface and its relaxation are limited practically to the 2-3 outmost surface layers. The directions of the magnetic moments do not change and the surface remains ferrimagnetic, but the magnitudes of the moments are smaller by about 0.1-1.6$\mu_{B}$ (GGA) and 0.05-0.56$\mu_B$ (GGA+$U$) compared to the corresponding atoms in magnetite bulk.  The strongest change of the magnetic moments is observed for the topmost surface layer. At the Fe$_{\rm tet1}$-termination the moments of the topmost Fe layer are reduced to 3.11$\mu_B$ (GGA) and 3.51$\mu_{B}$ (GGA+$U$). At the Fe$_{\rm oct2}$-terminated surface the moment is enhanced to 4.02$\mu_{B}$ (GGA) and 4.45$\mu_B$ (GGA+$U$).
As can be seen from the lowest panel of Fig.~\ref{relaxations}, this is connected with the big change in the relaxation of the topmost Fe$_{\rm oct2}$ layer which for the GGA+$U$ is practically reduced to zero. The atoms of subsurface O-planes of the Fe-terminated surfaces exhibit small magnetic moment $\approx$0.1-0.4$\mu_{B}$. 
The GGA+$U$ moments on oxygen atoms of the O$_1$ and O$_2$ terminated surfaces are in the range 0.15-0.23$\mu_{B}$ and ($-$0.02)-0.20$\mu_B$, for the O$_1$ and O$_2$ termination, respectively. For the O$_2$ termination one of the O atoms in the topmost layer has a very small ($\approx$0.02$\mu_{B}$) local moment opposite to the other three atoms.  The moments on the Fe atoms of subsurface planes in these two oxygen terminations show a significant reduction when calculated within GGA. In contrast, the GGA+$U$ values deviate relatively little from the moments on bulk atoms. 

\begin{figure}
\includegraphics[width=7.0cm]{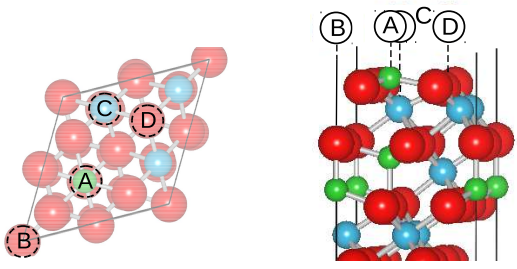}
\caption{(Color online) Schematic illustration of the initial Au and Pd adsorption places at the Fe$_{\rm tet1}$ terminated surface in a top (left) and side (right) view.} \label{fig_sites_Fetet1}
\end{figure}

\begin{figure}
\includegraphics[width=6.5cm]{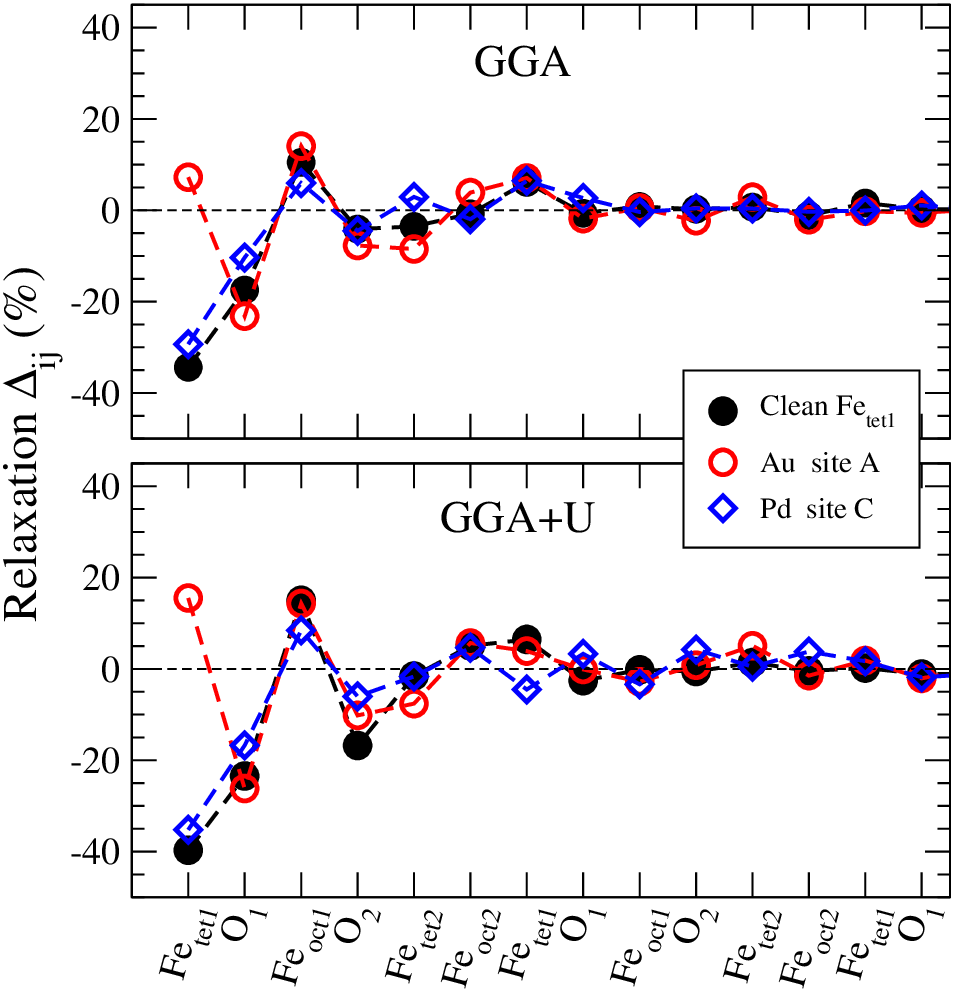}
\caption{(Color online) Relaxation of the interplanar distance after Au/Pd adsorption in most stable sites on the Fe$_{\rm tet1}$ terminated Fe$_3$O$_4$(111) surface. } \label{fig_relaxAuPd_Fe}
\end{figure}

\subsection{Adsorption on the Fe$_{\rm tet1}$ terminated surface}

The adsorbate atoms were initially placed well above four different sites  (Fig. \ref{fig_sites_Fetet1}) of the Fe$_{\rm tet1}$-terminated surface: in site A, on-top of the Fe$_{\rm tet1}$ atom, site D above the O$_1$ atom, or in one of the hollow sites, B or C, formed respectively in a deep hollow over the O$_2$ atom, and in the hollow above the Fe$_{\rm oct1}$ atom. All of the adsorption sites appeared to be stable after structure relaxations. The most stable position for Au adsorption is site A, whereas Pd atoms adsorb preferably in site C.

The changes in the surface relaxation due to Au/Pd adsorption are presented in Fig. \ref{fig_relaxAuPd_Fe}. The adsorption of Au and Pd in any considered adsorption site tends to suppress the surface relaxation of the first interplanar distance of the Fe$_{\rm tet1}$ termination compared to the clean surface. The relaxation patterns resulting from GGA and GGA+$U$ calculations are very similar, with the relaxation of the first four layers being larger for GGA+$U$. For Au adsorbed in the most preferred on-top site A, a large contraction of the clean surface (35\%, GGA;  40\%, GGA+$U$) is converted into distinct expansion ($\sim$10\% GGA; $\sim$15\%, GGA+$U$), which makes the separation larger than respective one in the bulk crystal. A Pd adatom in the on-top site A causes much smaller changes of the interplanar distance which remains contracted compared to that in bulk magnetite. In general, the changes in the surface geometry due to Pd adsorption are much smaller than those caused by Au. A Au/Pd atom adsorbed in any other considered site does not change qualitatively the relaxation pattern. For both adsorbates, the distances between deeper oxide layers show relatively small changes.

\begin{figure}
\includegraphics[width=7.0cm]{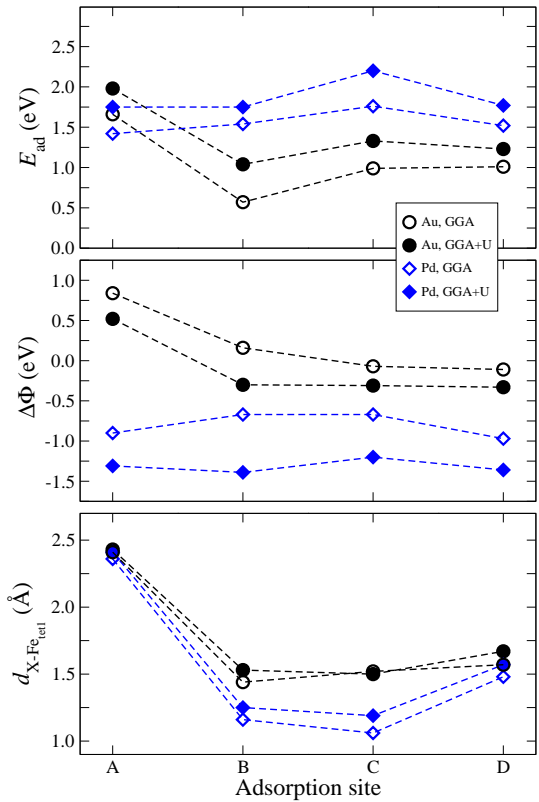}
\caption{(Color online) Adsorption energy, $E_{\rm ads}$,  work function change, $\Delta\Phi$, and adatom-surface distance for a single X (=Au,Pd) atom adsorbed in different sites of the Fe$_{\rm tet1}$ terminated surface.} \label{fig_adsAuPd_Fe}
\end{figure}

Figure \ref{fig_adsAuPd_Fe} displays the calculated adsorption energy and adatom-surface distance for a single Au/Pd atom adsorption. As is seen, in general, the GGA+$U$ binding is 0.2-0.5 eV stronger than that resulting from GGA. The Pd binding to the Fe$_{\rm tet1}$ terminated surface is stronger than that of Au in all sites, except for site A. This agrees with the results for Au/Pd adsorption on the iron terminated (0001) surface of hematite \cite{KiePab11}. The  adsorption energies in different sites are less differentiated for Pd than for Au. For Au adatoms the most stable position is site A (on top of the Fe$_{\rm tet1}$ atom), whereas for Pd it is the threefold O-coordinated hollow C (Fig.~\ref{fig_sites_Fetet1}). The binding energy of the Au (Table \ref{tab2_ads_Fetet1}) in the most preferred on-top site A is 1.66 eV (GGA) and 1.98 eV (GGA+$U$). In this site, the Au--Fe$_{\rm tet1}$ bond is perpendicular to the magnetite surface with a bond length of 2.43 \AA\ (GGA and GGA+$U$).
For Pd adsorption the most stable site is the threefold O-coordinated hollow site C with an adsorption energy of 1.76 eV (GGA) and 2.20 eV (GGA+$U$). The respective Pd-O bond lengths are 2.18 {\AA}, 2.46 {\AA}, and 2.46 {\AA}, which  is not much larger (2.68 \AA) than the distance between the Pd and Fe$_{\rm oct1}$ atom in the third surface layer. 

\begin{table}
\caption{Adsorption energy, $E_{\rm ads}$, adatom-surface distance, $d_{\rm X-Fe_{tet1}}$, and work function change, $\Delta\Phi$, for X (=Au,Pd) adatom in most stable sites on the Fe$_{\rm tet1}$-terminated surface. For each quantity the left and right hand side columns display GGA and GGA+$U$ values, respectively.} \label{tab2_ads_Fetet1}
\begin{ruledtabular}
\begin{tabular}{c|cc|cc|cc}
X (site) & \multicolumn{2}{c|}{$E_{\rm ads}$ (eV)} & \multicolumn{2}{c|}{$d_{\rm X-Fe_{tet1}}$ ({\AA})} & \multicolumn{2}{c}{$\Delta\Phi$ (eV)} \\
\hline
Au (A) & 1.66 & 1.98 & 2.41 & 2.43 & 0.84 & 0.52\\
Pd (C) & 1.76 & 2.20 & 1.06 & 1.19 & -0.67 & -1.20\\ 
\end{tabular}
\end{ruledtabular}
\end{table}

\begin{figure*}
\includegraphics[width=15cm]{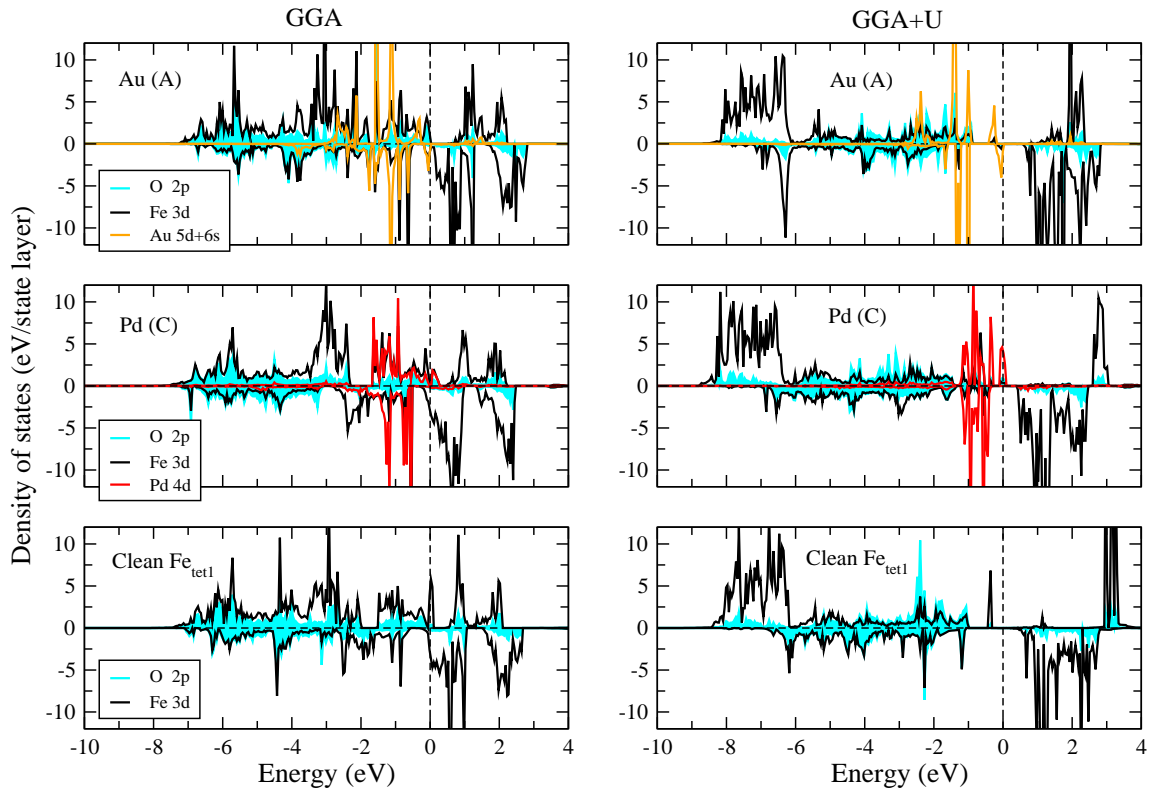}
\caption{(Color online) Local density of states for Au and Pd adsorption in most stable sites on the Fe$_{\rm tet1}$ termination resulting from GGA and GGA+$U$ calculations. LDOS of the adsorbate atom and of the three (two Fe and one O) topmost atomic layers of the magnetite (111) surface are shown. Corresponding LDOS for the clean Fe$_{\rm tet1}$-termination are displayed for comparison}
\label{ldos_AuPd_Fetet1}
\end{figure*}

The stronger binding of Pd than Au can be understood by inspecting the results of the layer-resolved LDOS presented in Fig.~\ref{ldos_AuPd_Fetet1}. They were calculated for the Au in site A and the Pd in site C, which are their most stable positions. The GGA LDOS of surface layers with Au and Pd show metallic character. The principal peaks of the Pd $4d$ electron states are higher in the energy and hybridize with the O $2p$ states closer to the Fermi level than the Au $5d$ peaks, which results in stronger binding of the Pd. In the case of Au adsorption the metallicity of the oxide is enhanced by the Au $5d$ states in the energy range between -3 eV and the Fermi level and a smaller contribution from the Au $6s$ states  at the energies close to the Fermi level. In the GGA+$U$ approach the adsorbed Au introduces a small density of $6s$ states at the Fermi level which contributes to the reactivity of this surface and changes the semiconducting character of the surface. The Pd $4d$ states tend to close the energy gap on this oxide termination and convert it from semiconductor to semi-metal. For the GGA+$U$ the main weight of the Pd $4d$ states is about 1 eV closer to the Fermi level than that of the Au $5d$ states, which again means stronger binding of Pd than Au. 

The differences between Au and Pd adsorption are also seen in the work function changes (Fig.\ \ref{fig_adsAuPd_Fe}).  In all considered sites Pd lowers the work function while  Au distinctly increases it only when adsorbed in the A site. In the remaining sites Au either slightly reduces the work function (GGA+$U$) or has only a very small effect on it (GGA). In general, GGA predicts a higher work function increase (by $\approx$0.3 eV) than GGA+$U$ for Au adsorption and a smaller (by 0.4-0.7 eV) lowering of the work function due to Pd adatom. For Au adsorption in the most preferred site A, the work function is increased by 0.84 eV (GGA) and 0.52 eV (GGA+$U$). In contrast, the work function of the Pd/Fe$_{\rm tet1}$ system, with Pd in the most stable site C, is lower by 0.67 eV (GGA) and 1.20 eV (GGA+$U$) compared with that of the clean surface.
The above work function changes are consistent with the  electron charge transfer to/from the surface atoms due to adsorption of Au/Pd. The Bader charge analysis \cite{Bader90,HenAJ06} shows that a Au adatom in site A gains electrons (0.27$e$, GGA; 0.32$e$, GGA+$U$) at the expense of the surface Fe and O atoms. In the case of  GGA+$U$ most of the charge (0.22$e$) is donated by the Fe$_{\rm tet1}$ atom from beneath the Au. In the case of Pd adsorption in the site C, the charge (0.42$e$, GGA; 0.36$e$, GGA+$U$) is transferred from the adatom to the surface, mostly to the Fe$_{\rm tet1}$ atom (0.13$e$, GGA; 0.05$e$, GGA+$U$).

The changes of the magnetic moments caused by Au/Pd adsorption are small and limited to substrate atoms in the  2-3 topmost layers. For Au in the most stable site A, the GGA+$U$ magnetic moment on the topmost Fe$_{tet1}$ atoms is enhanced by about 0.42$\mu_{B}$ with respect to that of the clean surface. For Pd GGA+$U$ adsorption in the most preferred site C the change is negligible ($\approx$0.01$\mu_{B}$).  
Small magnetic moments appear on the Au and Pd adatoms. The moment on the Au atom in different sites is in the range of 0.14-0.20$\mu_B$ (GGA) and 0.10-0.19$\mu_{B}$ (GGA+$U$), the lowest values corresponding to the Au atom in most stable site A, and has the same direction as the moments on the Fe atoms in the topmost Fe$_{tet1}$ layer. The magnetic moment on the Pd adatom in sites A, C, and D is in the range of 0.07-0.27$\mu_{B}$ (GGA), and 0.03-0.16$\mu_{B}$ (GGA+$U$), and has the same direction as the moment on the Fe atoms in the topmost surface layer. The magnetic moment on Pd at the site B is 0.41$\mu_{B}$ and 0.62$\mu_{B}$ for GGA and GGA+$U$, respectively, and is oriented upwards. The moment on the Pd adatom in the most preferred site C is the same (-0.09$\mu_{B}$) for GGA and GGA+$U$ calculations.  

\begin{figure}
\includegraphics[width=7.0cm]{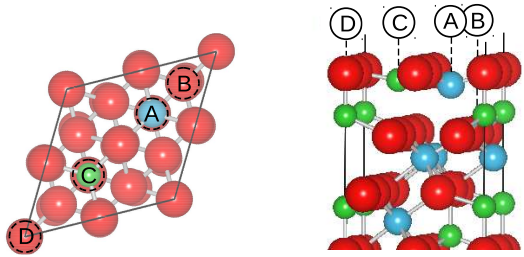}
\caption{(Color online) Schematic illustration of the initial Au and Pd adsorption places at the O$_2$-terminated Fe$_3$O$_4$(111) surface in a top and side view.}\label{fig_sites_O2}
\end{figure}

\subsection{Adsorption on the O$_2$ terminated surface}

For the oxygen termination the following four different adsorption sites  (Fig.~\ref{fig_sites_O2}) were considered: the hollow over the Fe$_{\rm oct2}$ atom (site A), a deep hollow over the O$_1$ atom (site B), a hollow over the Fe$_{\rm tet2}$ atom (site C), and the site above (on-top) the lower O$_2$ (site D). All of the considered positions appeared to be stable during structure optimization, and the energetically preferred site, both for Au and Pd adsorption, is site B, where the adsorbate atom is coordinated by three O atoms.  

\begin{figure}
\includegraphics[width=6.5cm]{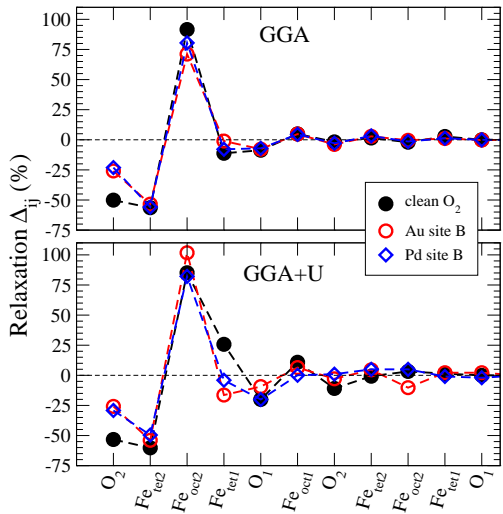}
\caption{(Color online) Interlayer relaxation (in \%) of the interplanar distance after Au and Pd adsorption in most stable sites on the O$_2$ terminated Fe$_3$O$_4$(111) surface.} \label{fig_relax_AuPd_O2}
\end{figure}

The adsorbate induced changes in the relaxation pattern resulting from GGA and GGA+$U$ calculations are qualitatively and quantitatively similar for both types of adsorbates (Fig.~\ref{fig_relax_AuPd_O2}). The adsorption of a Au/Pd atom on the O$_2$ terminated surface suppresses the relaxations of the first two interplanar distances. The contraction of the first interplanar distance is reduced by one-half with respect to the clean surface, but the distance still remains 25\% shorter than that in the bulk crystal. The changes of the relaxations of deeper distances are rather small. Interestingly, the difference in the sign of the relaxation of the fourth interlayer distance in the GGA and GGA+$U$ results is removed in the adsorbate--oxide system.

\begin{figure}
\includegraphics[width=7.0cm]{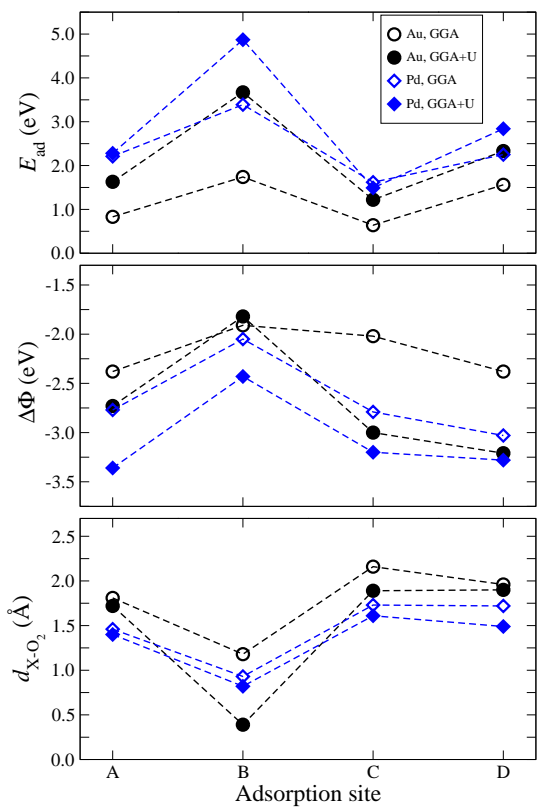}
\caption{(Color online) The same as in Fig. \ref{fig_adsAuPd_Fe} but for the O$_2$ terminated  surface.}\label{fig_ads_AuPd_O2}
\end{figure}

The adsorption binding energy in the considered sites and the adatom distance to the topmost oxide layer are displayed in Fig.~\ref{fig_ads_AuPd_O2}. Additionally, Table \ref{tab3_ads_O2} shows numerical values for the most stable site.  In general, the GGA+$U$ binding is up to about 2 eV larger than that calculated within GGA. Similarly as for the Fe$_{\rm tet1}$ termination the Pd atoms bind stronger than Au to the O$2$-terminated surface. The Au binding energy is 1.74 eV (GGA) and 3.66 eV (GGA+$U$), whereas for Pd it is 3.39 eV (GGA) and 4.87 eV (GGA+$U$). The bond lengths between Au and the three coordinating oxygens are not symmetric and are 1.99, 2.07, and 2.07 {\AA} for GGA+$U$ (2.03, 2.29 and 2.29 {\AA} for GGA). These numbers  are very close to the corresponding Pd-O bond lengths of 1.95, 2.10, and 2.10 {\AA}.  

\begin{table}
\caption{Adsorption energy, $E_{\rm ads}$, adatom-surface distance, $d_{\rm X-O_2}$, and work function change, $\Delta\Phi$, for X (=Au,Pd) adatom in most stable sites on the O$_{2}$-terminated surface. For each quantity the left and right hand side columns display GGA and GGA+$U$ values, respectively.} \label{tab3_ads_O2}
\begin{ruledtabular}
\begin{tabular}{c|cc|cc|cc}
 X (site) &\multicolumn{2}{c|}{$E_{\rm ads}$ (eV)} & \multicolumn{2}{c|}{$d_{\rm X-O_2}$ ({\AA})} & \multicolumn{2}{c}{$\Delta\Phi$ (eV)}\\
\hline
Au (B) & 1.74 & 3.67 & 1.18 & 0.39 & -1.91 & -1.82\\
Pd (B) & 3.39 & 4.87 & 0.93 & 0.82 & -2.05 & -2.43\\ 
\end{tabular}
\end{ruledtabular}
\end{table}

\begin{figure*}
\includegraphics[width=15cm]{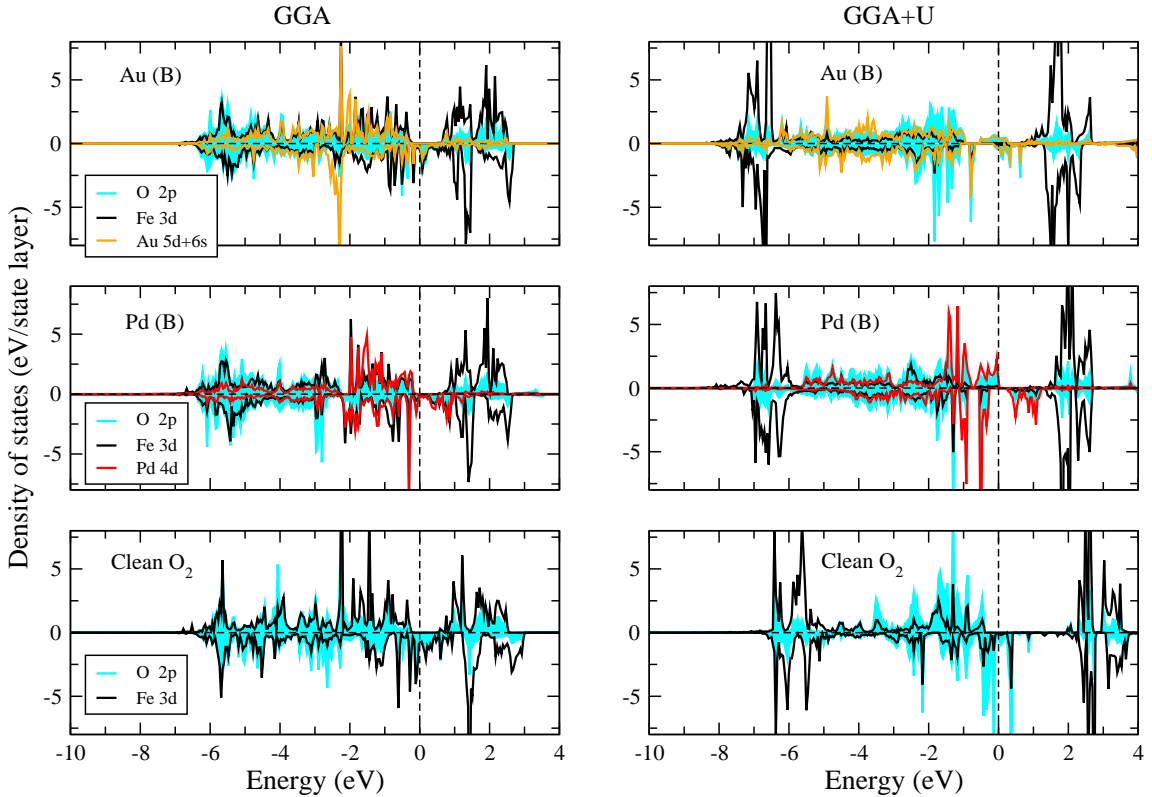}
\caption{(Color online) Same as in Fig.~\ref{ldos_AuPd_Fetet1} but for Au/Pd adsorption on the O$_2$ termination. }
\label{ldos_AuPd_O2}
\end{figure*}

Figure \ref{ldos_AuPd_O2} displays the layer-decomposed LDOS for the Au and Pd adsorbed in the energetically favored site B. The adsorbate covered surface remains half-metallic both when calculated in GGA and GGA+$U$ although in the latter case a nonvanishing LDOS of the hybridized Au $5d$ and O $2p$ states at the Fermi level makes the bands look more metallic-like. The Au and Pd states are much more delocalized and extended over a wider range of energies compared with the Fe$_{\rm tet1}$ termination. The half-metallicity of the GGA+$U$ bands of the Au/O$_2$ system is due to minority spin band of the Fe $3d$ states. The energy gap in the majority spin band is shifted to lower energies compared to the clean surface.  
Upon Au adsorption the hybridization of the Au 5$d$ and O $2p$ states is strongest up to about 0.5 eV below the Fermi level while for the Pd $4d$ states the hybridization with the O $2p$ states occurs up to the Fermi level. This stronger  (compared to that of Au) hybridization of the oxide and Pd states at energies closer to the Fermi level explains the stronger binding of Pd than Au to the O$_2$ terminated surface (Fig.~\ref{fig_ads_AuPd_O2}).
The resulting LDOS are dominated by the O $2p$  electron states of the surface O$_2$ layer hybridized with the Au 5$d$  or Pd $4d$ states. The states of Fe atoms of the underlying Fe$_{\rm tet2}$ and Fe$_{\rm oct1}$ layer contribute significantly only to the LDOS at the lower energy range, below $-6$ eV.
For Pd adsorption the LDOS on the Pd atom is more localized than on Au and its main weight extends between $-1.5$ eV and the Fermi level. In contrast to the clean termination, in the presence of adsorbed Pd, responsible for half-metallic character of the surface is the energy gap in the minority spin band.

The work function of the oxygen terminated surface decreases dramatically upon Au/Pd adsorption (Fig.~\ref{fig_ads_AuPd_O2}). Au adsorption in the most preferred site B decreases the work function by about 1.9 eV (GGA) and 1.73 eV (GGA+$U$). For Pd adsorption this decrease is even larger: 2.0 eV (GGA) and 2.4 eV (GGA+$U$) which means a nearly 30\% reduction with respect to the clean surface value. A reduction of work function is connected with a decrease of the surface dipole moment and indicates a decrease in polarity. 
Both for Au and Pd a charge transfer from the adsorbate to the substrate is obtained. This is different from adsorption on the iron terminated surface, where a transfer from adsorbate to the surface is observed only for Pd, but is consistent with the large work function decrease described above. The Au adatom loses 0.70$e$ (GGA) or 0.77$e$ (GGA+$U$), Pd loses 0.89$e$ (GGA) or 0.92$e$ (GGA+$U$). Most of this charge is transferred to the atoms of the O$_2$ layer closest to the adatom (0.08$e$ (GGA) and 0.30$e$ (GGA+$U$) for Au, and 0.28$e$ (GGA) and 0.25$e$ (GGA+$U$) for Pd adsorption).   

The magnetic moments on the oxygen atoms in the topmost surface layer are only little affected by the presence of adsorbate. For a Au atom in the most preferred site B, the GGA+$U$ moments on the O-atoms are changed by 0.11-0.16$\mu_{B}$. A Pd atom in the same site induces the smaller changes (0.02-0.07$\mu _B$). For both adsorbates in the most stable site B, the moment of one of the O atoms in the topmost substrate layer is oriented down which is opposite to the direction of the remaining three O atoms, and is similar to the clean O$_2$ terminated surface. 
The changes on the Fe$_{\rm tet2}$ atoms of the first subsurface layer resulting from the GGA+$U$ calculations for Au and Pd adsorption are -0.40$\mu_{B}$ and -0.29$\mu_{B}$,  respectively. A decreased magnitude of the magnetic moments of the Fe$_{\rm tet2}$ atoms makes them comparable with those in the bulk magnetite. The changes of moments of atoms in deeper layers compared with the bulk are negligible.  
On the O$_2$-terminated surface the magnetic moments on the Au differ from those on Pd atoms. The moments of the Au atom are lower than 0.1$\mu_{B}$. Only the magnetic moments of the Au atom in the site B and C resulting from the GGA are larger than that (0.13$\mu_B$ and $-$0.15$\mu_B$, respectively). The moment on Au in the most preferred site B is $-$0.02$\mu_{B}$ (GGA+$U$). The magnetic moments on Pd are much larger than those on the Au and are in the range 0.4-0.9$\mu_{B}$. In the most stable site B the magnetic moment on Pd is 0.81$\mu_{B}$ (GGA) and 0.92$\mu_{B}$ (GGA+$U$). 

\section{Summary and conclusion}
We have presented a detailed DFT and DFT+$U$ study of the structural, electronic, and magnetic properties of the clean magnetite (111) surface, and the adsorption of Au and Pd atoms on its two stable, Fe$_{\rm tet1}$ and O$_2$, terminations. Inclusion of on-site Coulomb correlations in GGA+$U$ approach modifies profoundly the electronic structure of magnetite surfaces. Based on \textit{ab initio} thermodynamics the Fe$_{\rm tet1}$ terminated surface is confirmed to be the most stable one over a broad range of oxygen pressures. It shows metallic character when calculated within GGA and half-metallic DOS with zeroth LDOS at the Fermi level when calculated within GGA+$U$. All terminations studied exhibit a large inward relaxation of the first interlayer distance. Both Au and Pd bind strongly to the magnetite surface and induce large changes in the surface geometry. At the Fe$_{\rm tet1}$ termination different sites are favored for adsorption of Au and Pd. For Au adsorption the most favorable site is on top of the Fe$_{\rm tet1}$ atom, whereas for Pd it is threefold coordinated hollow. At the O$_2$ termination the threefold hollow site B, where the adatom is coordinated by three O atoms, is most stable both for Au and Pd adsorption. 
The GGA+$U$ bonding is respectively 0.2-0.5 eV and 0.5-1.5 eV stronger on the iron and the oxygen terminated surface, than that resulting from standard GGA calculations. The binding is stronger for Pd than Au and for both adsorbates is distinctly stronger on the oxygen than on the iron terminated surface. 
The Au/Pd bonding either to iron or oxygen terminated (111) surface of magnetite shows close resemblance to that reported by us for Au/Pd adsorption on the hematite (0001) surface terminations both with respect to the preference of the sites and strength of  the bonding \cite{KiePab11}. 

\begin{acknowledgments} 
This work was supported by the Polish Ministry of Science and Higher Education in the years 2008-11 under Grant No.\ N N202 072535. We are are grateful to Ernst Bauer for useful comments. AK acknowledges hospitality of Risto Nieminen, Aalto University, Finland, and the access to high performance computers of the CSC Espoo, Finland, under the HPC-Europa2 project (No.\ 228398) with the support of the European Commission -- Capacities Area -- Research Infrastructures. We also acknowledge provision of computer time from the Interdisciplinary Centre for Mathematical and Computational Modelling (ICM) of the Warsaw University within the Project No.\ G44-23.
\end{acknowledgments}


\end{document}